\documentclass[a4paper,fleqn,usenatbib]{mnras}
\usepackage[T1]{fontenc}
\usepackage{ae,aecompl}
\usepackage{graphicx}	% Including figure files
\usepackage{amsmath}	% Advanced maths commands
\usepackage{amssymb}	% Extra maths symbols

\title[Clumpy structure in NGC~2548]{NGC~2548: clumpy spatial and 
kinematic structure in an intermediate-age galactic cluster}

\author[Vicente et al.]{
Bel\'en Vicente$^1$,
N\'estor S\'anchez$^{1,2}$\thanks{E-mail: nestor@um.es (NS)}
and Emilio J. Alfaro$^{1}$\\
$^{1}$Instituto de Astrof\'{\i}sica de Andaluc\'{\i}a, CSIC,
Glorieta de la Astronom\'{\i}a s/n, 18008, Granada, Spain\\
$^{2}$Departamento de F\'{\i}sica, Universidad de Murcia,
E-30100 Murcia, Spain.}

\date{Accepted XXX. Received YYY; in original form ZZZ}
\pubyear{2016}

\begin{document}
\label{firstpage}
\pagerange{\pageref{firstpage}--\pageref{lastpage}}
\maketitle

\begin{abstract}
NGC~2548 is a $\sim400-500$~Myr old open cluster with evidence of
spatial substructures likely caused by its interaction with the
Galactic disk. In this work we use precise astrometric data from
the {\it Carte du Ciel} - San Fernando (CdC-SF) catalogue to
study the clumpy structure in this cluster. We confirm the 
fragmented structure of NGC~2548 but, additionally,
the relatively high precision of our kinematic data
lead us to the first detection of substructures in the proper
motion space of a stellar cluster. There are three spatially
separated cores each of which has its own counterpart in the proper
motion distribution. The two main cores lie nearly parallel to the
Galactic plane whereas the third one is significantly fainter than the
others and it moves toward the Galactic plane separating from the
rest of the cluster. We derive core positions and proper motions,
as well as the stars belonging to each core.
\end{abstract}


\section{Introduction}

It is generally accepted that most stars (between $70-90$\%) are
born in star clusters \citep{Lad03,Pis08}, which are naturally
associated to high density regions in molecular clouds
\citep{Elm96,Elm06}. Few million years after its formation,
the number of observed clusters falls drastically
\citep{Oor58,Wie71,Lad10}. A rapid disruption of these stellar
systems, which affects a high proportion of them, occurs at early
dynamical evolutionary stages when embedded clusters remove their
gas component \citep{Tut78}.
At this time, stellar winds from massive stars and
ejected matter from supernova explosions dissipate parental
intra-cluster gas. If kinetic energy of the stars exceeds the
remaining gravitational potential then the cluster quickly
dilutes the primordial assembly into the stellar galactic field.
Only $\sim10$\% of the number of original clusters seem to survive
the so-called ``infant mortality" and remain bound for longer
than $1$~Gyr \citep{Fal05}. Surviving stellar clusters are suffering
processes of partial destruction and dissipation during their whole
life, whose true nature and details are still matter of debate
although several global scenarios have been proposed in the last
years \citep[e.g.][]{Elm10}. Tidal galactic field and encounters
with giant molecular clouds are two main mechanisms responsible
for the dissipation and disruption of stellar clusters, but the
total lifetime of these systems and the rate at which they are
loosing stars are very dependent on cluster initial conditions
\citep{Elm10}
and on the main destruction mechanism; for example, it
has been shown that a star cluster can be disrupted even by a 
single tidal encounter with a giant molecular cloud \citep{Kru11}.
Thus, initial mass, spatial distribution of zero-age stars,
galactic location and local environment of the birth place
affect the dynamical evolution of stellar clusters for the
first several hundred million years \citep{Par14a,Par14b}.

Observations indicate that most embedded clusters, as well as
some intermediate-age systems which have already lost their
parental gas component, may exhibit a clumpy structure formed
by several connected blobs or filaments \citep{Lad03}. The
spatial distribution of stars in these clusters can be
described as a fractal pattern and quantified by a single
parameter \citep[e.g.][]{Car04,San09}. It seems that the
original clumpy structure in some clusters tends to be erased
with age giving rise to radial density profiles, although
clusters as old as $\sim100$~Myr may still present fractal
patterns \citep{San09}.
However, there exist a number of different
possible morphologies and alternative scenarios have
been proposed in which structurally simple clusters
are built up through hierarchical merging of smaller
subclusters \citep{Kuh14,Kuh15}. Thus, a
key question is whether the clumpy internal
structure observed in some clusters is just the
imprint of the parental gas cloud, a signature that tends to
disappear over time, or whether new blobs can be generated
and/or maintained during the cluster dynamical evolution.
A suitable strategy to address this problem is to find clusters
having evidence of internal clumpiness and then collect and analyse
in detail the spatial and kinematic information available for them.
Galactic open clusters showing signatures of spatial substructure
have been previously detected and analysed. Most examples refer to
young ($\lesssim5$~Myr) clusters whose structure can be mainly
attributed to their original clouds \citep{Lad03,Gre15}, for
instance, NGC~1333 \citep{Lad96} and NGC~2264 \citep{Fur06}.
Intermediate age (a few times $100$~Myr) open clusters with clear
evidence of spatial substructures include NGC~2287, NGC~2516 and
NGC~2548 \citep{Ber01}. Irregular patterns have been also observed
in old ($\gtrsim 1$~Gyr) clusters like M67 \citep{Dav10} and
recently in NGC~6791 \citep{Dal15}. 
The irregular overdensities
observed in intermediate-age and old clusters are interpreted as
associated with real structures stretched by the Galactic potential
field \citep{Ber01}, but the details of this mechanism can only be
fully evaluated through an accurate analysis of the kinematics of
cluster members on a large spatial scale to compare core and halo 
dynamics. The search of patterns or substructures in the velocity
phase space requires further improvement both in specific searching
tools \citep[e.g.][]{Alf16} and in the precision of the kinematic
data. Precise radial velocity measurements have allowed to resolve
clumpy structures in four clusters: NGC~2264 \citep{Fur06,Tob15},
Orion Nebula Cluster \citep{Fur08}, Gamma Velorum \citep{Jef14}
and NGC~2547 \citep{Sac15}. There is not, to our knowledge, any
reported evidence of grouping or patterns in the proper motion
subspace.

From the above mentioned clusters, NGC~2548 is contained in the
{\it Carte du Ciel} - San Fernando (CdC-SF)
astrometric catalogue \citep{Vic10}. The CdC-SF catalogue
presents some advantages over proper motion data currently
available for this cluster. It contains precise measurements
($\sim2$~mas~yr$^{-1}$ for stars to V$\sim$16 and $\sim1.2$~mas~yr$^{-1}$
for V$\sim$14) which means a deeper extension of Hipparcos, in
terms of proper motions, up to a magnitude of 14. In this work
we use these data to study the properties of NGC~2548, both in
spatial coordinates and in proper-motion space.
NGC~2548 (M48) is a well known open cluster with many studies
since the second half of XX century providing extensive knowledge
of its main physical properties. It is located in the third
galactic quadrant with coordinates $\alpha = 08$h $13$m $43$s
and $\delta = -05\degr$ $45\arcmin$ \citep{Dia02}. The latest
studies generally agree on a distance value in the range
$700-780$~pc, a very low reddening $E(B-V)=0-0.1$ and nearly
solar metallicity \citep{Wu02,Rid04,Bal05,Sha06,Wu06,Bar15}.
The most recent age estimation, based on stellar rotation
periods, yielded $450\pm50$~Myr \citep{Bar15} in well agreement
with previously determined isochrone ages \citep{Rid04,Bal05}.
This cluster is
specially suitable for dynamical evolution studies because of
its internal clumpy structure formed by three well differentiated
blobs \citep[see Fig.~5 in][]{Ber01}, but the available kinematic
data suffer from one or more of the following drawbacks: (a) small
number of data points, (b) limited to relatively bright stars,
(c) partial covering of the cluster area and (d) large proper
motion uncertainties. By using data from the CdC-SF catalogue,
we are able to overcome these issues and to carry out a detailed
analysis of the internal structure of NGC~2548.
This paper is organized as follows. Section~\ref{sec:sample}
describes the observational data used in this work. Cluster
memberships are derived in Section~\ref{sec:members} and then
used to calculate the spatial and kinematic distribution of
stars in Section~\ref{sec:maps}. These results are discussed
in Section~\ref{sec:discussion} in the context of the dynamical
evolution of star clusters. Finally, the main conclusions are
summarised in Section~\ref{sec:conclusion}.

\section{Sample of stars}
\label{sec:sample}

\begin{figure}
\includegraphics[width=\columnwidth]{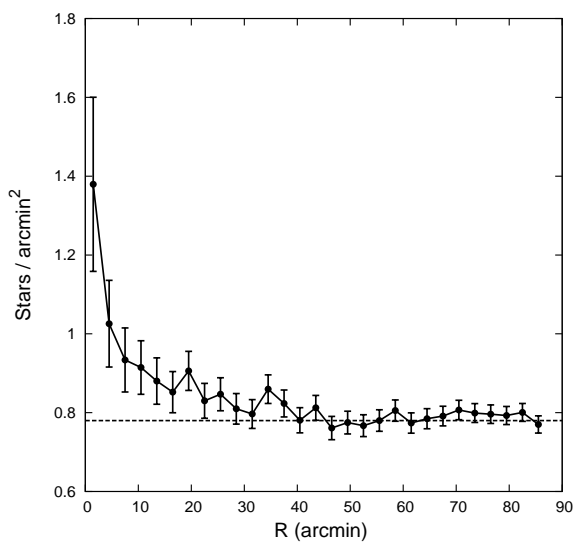}
\caption{Radial density profile of stars toward
NGC~2548. Error bars are from Poisson statistics.
Horizontal dashed line indicates the mean
value of $0.78$ stars~arcmin$^{-2}$ calculated
from $R=45$~arcmin, that we consider as optimal 
sampling radius.}
\label{fig:perfil}
\end{figure}

We use the CdC-SF astrometric catalogue \citep{Vic10} which
provides precise positions and proper motions for stars up
to a magnitude of V$\sim$16. The mean positional uncertainty
is $0.20$~arcsec ($0.12$~arcsec for well-measured stars) and
the proper motion uncertainty is $2.0$~mas~yr$^{-1}$
($1.2$~mas~yr$^{-1}$ for well-measured stars). For extracting
the data from this catalogue we choose a circle centred on
the cluster coordinates given by \citet{Dia02},
$\alpha = 123.43$~deg and $\delta = -5.75$~deg,
that we will assume as the cluster centre in this work.
As discussed in \citet{San10}, a proper choice of the
sampling radius is crucial to avoid some biases and
problems determining memberships and, therefore, the
remaining cluster properties. We tried to follow the
recipe suggested in \citet{San10}, based on analysing
memberships for several sample radii, but given the
irregular space and velocity distributions of stars
in this cluster (see next sections) this has not been
possible. In any case, the optimal sampling radius
nearly coincides with the cluster radius itself
\citep{San10}; however, it is rather unreliable
to estimate this radius from the cluster members
available in the literature because, independently
of the used method, assigned members tend to spread
throughout the area selected by the author. This
fact produces discrepancies in the derived radius
values.
From the projected radial density of stars it is
possible to estimate the ``tidal" radius as the
point where the radial density becomes roughly
constant and merges with the field star density.
In this way, tidal radius values ranging from
$\sim 8$~arcmin \citep{Sha06} to $\sim 43.8$~arcmin
\citep{Kar05} have been reported. We have calculated
the surface density of stars (Fig.~\ref{fig:perfil})
from which we obtain $R=45$~arcmin as the optimal
sampling radius. With this radius we extract positions
and proper motions for the stars in the CdC-SF catalogue
in the direction centred at the NGC~2548 coordinates. The
total number of stars in our sample is $1655$.

%%%%%%%%%%%%%%%%%%%%%%%%%%%%%%%%%%%%%%%%%%%%%%%%%%%%%%%%%%%%
%%%%%%%%%%%%%%%%%%%%%%%%%%%%%%%%%%%%%%%%%%%%%%%%%%%%%%%%%%%%

%%%%%%%%%%%%%%%%%%%%%%%%%%%%%%%%%%%%%%%%%%%%%%%%%%%%%%%%%%%%
\begin{table*}
\caption{List of the $1655$ stars toward NGC~2548.
This is only a portion for guidance regarding its
form and content. The full table is available online.}
\label{tab:main}
\begin{tabular}{cccccccc}
\hline
UCAC4-ID & $\alpha$ (deg) & $\delta$ (deg) & 
$\mu_\alpha\cos\delta$ (mas~yr$^{-1}$) & $\mu_\delta$ (mas~yr$^{-1}$) &
prob. & $B$ (UCAC4) & $V$ (UCAC4) \\
(col 1) & (col 2) & (col 3) & (col 4 $\pm$ 5) & (col 6 $\pm$ 7) &
(col 8) & (col 9 $\pm$ 10) & (col 11 $\pm$ 12) \\
\hline
%  UCAC4  RAdeg  DECdeg  muRAcosD  e_muRA  muDEC  
%  e_muDEC  prob  Bmag  e_Bmag  Vmag  e_Vmag
$421-044734$ & $122.6813$ & $-5.8258$ &  $-3.44\pm0.17$  &  
$+2.07\pm0.24$ & $0.92$ & $13.99\pm0.02$ & $13.56\pm0.01$ \\
$422-043693$ & $122.6840$ & $-5.6603$ & $-14.10\pm10.02$ & 
$+15.65\pm1.52$ & $0.00$ &                &              \\
$421-044741$ & $122.6892$ & $-5.8644$ &  $+0.65\pm0.40$  &  
$-1.98\pm0.24$ & $0.84$ & $13.83\pm0.01$ & $13.31\pm0.02$ \\
$422-043697$ & $122.6908$ & $-5.6704$ &  $-4.02\pm1.24$  &  
$+0.98\pm1.80$ & $0.89$ & $14.87\pm0.02$ & $14.19\pm0.01$ \\
$422-043698$ & $122.6916$ & $-5.6399$ &  $-6.96\pm1.52$  &  
$+6.29\pm1.71$ & $0.79$ & $14.99\pm0.03$ & $14.40\pm0.03$ \\
$421-044744$ & $122.6929$ & $-5.8273$ &  $-5.01\pm0.37$  &  
$-2.22\pm0.35$ & $0.55$ & $10.59\pm0.08$ & $9.75\pm0.10$ \\
$421-044745$ & $122.6931$ & $-5.8574$ &  $-3.89\pm0.32$  &  
$+4.65\pm0.32$ & $0.93$ & $13.28\pm0.01$ & $12.28\pm0.00$ \\
$423-045526$ & $122.6977$ & $-5.5697$ &  $-1.09\pm0.40$  &  
$+5.67\pm1.24$ & $0.93$ & $13.59\pm0.04$ & $12.56\pm0.01$ \\
$421-044751$ & $122.6983$ & $-5.8903$ &  $+1.95\pm1.14$  &  
$-3.54\pm1.71$ & $0.68$ & $14.75\pm0.03$ & $14.03\pm0.03$ \\
$422-043700$ & $122.6997$ & $-5.7492$ &  $-4.97\pm0.58$  &  
$-2.97\pm0.32$ & $0.43$ & $14.14\pm0.03$ & $13.59\pm0.02$ \\
\hline
\end{tabular}
\end{table*}
%%%%%%%%%%%%%%%%%%%%%%%%%%%%%%%%%%%%%%%%%%%%%%%%%%%%%%%%%%%%

%%%%%%%%%%%%%%%%%%%%%%%%%%%%%%%%%%%%%%%%%%%%%%%%%%%%%%%%%%%%
\begin{figure*}
\includegraphics[width=\columnwidth]{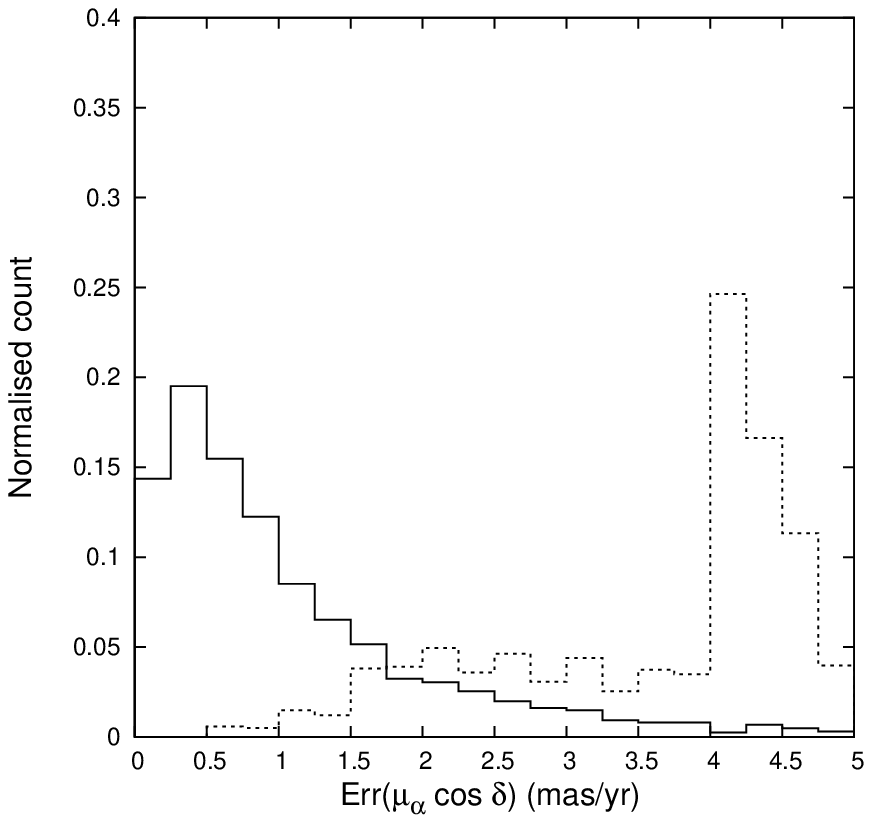}
\includegraphics[width=\columnwidth]{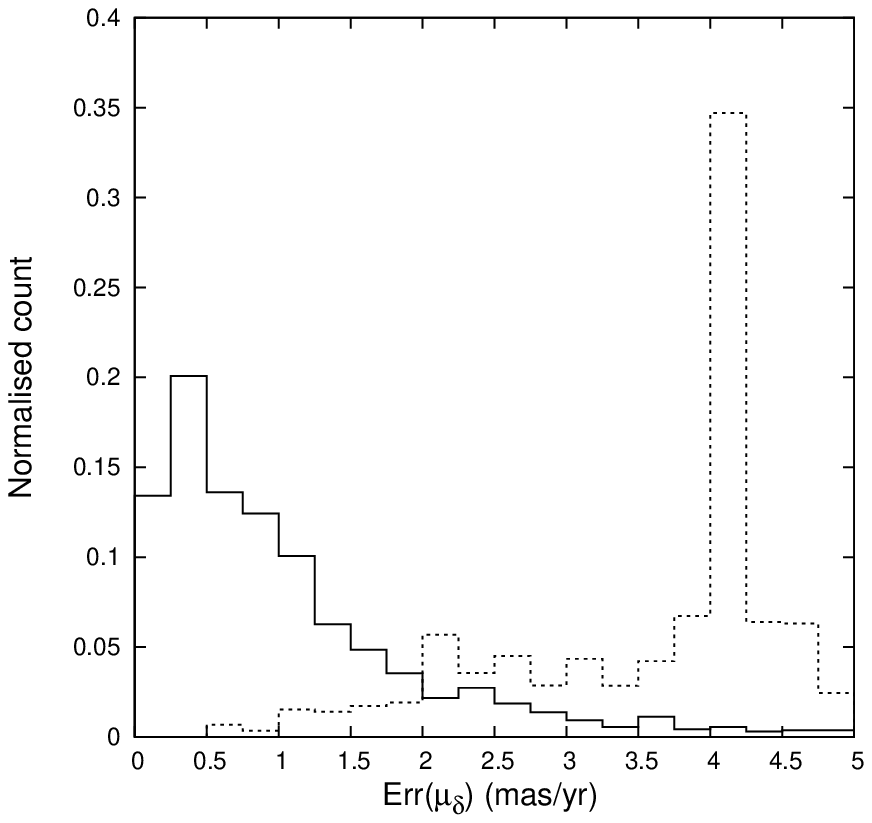}
\caption{Distribution of proper motion errors in right
ascension (left panel) and declination (right panel) for
all the stars in our sample. Solid lines correspond to
data from the CdC-SF catalogue whereas dashed lines
refer to UCAC4.}
\label{fig:errores}
\end{figure*}
%%%%%%%%%%%%%%%%%%%%%%%%%%%%%%%%%%%%%%%%%%%%%%%%%%%%%%%%%%%%

From the previous studies of NGC~2548 with photometric
data available online \citep{Rid04,Bal05,Wu06,Bar15},
we note that either the sample of stars is too small
for our purposes or the studied area does not fully
cover the area we are analysing here. What we have
done is to complete our spatial and kinematic information
with photometric data from the UCAC4 catalogue \citep{Zac13},
which have photometry for $\sim113$ millions of stars covering
the entire sky in the $BgVri$ bands supplemented with $JHK$
from 2MASS \citep{Skr06}.
Both catalogues were matched by moving CdC-SF positions
from 1900 to the UCAC4 mean epoch 2000 and then finding
the nearest star in UCAC4 for each star in CdC-SF.
Position differences were smaller than 0.2 arcsec
(the CdC-SF mean positional uncertainty) for almost
all stars. For the cases for which separations were
$\ga 0.2$ arcsec we additionally verified that
proper motions from CdC-SF and UCAC4 did not 
differ within their respective uncertainties.
Fig.~\ref{fig:errores} shows the errors in proper
motion for our sample of stars compared with the
errors extracted from the UCAC4 catalogue. The mean
value of our errors is $\sim 1.0$~mas~yr$^{-1}$ and
the median is $\sim 0.7$~mas~yr$^{-1}$. The peak of the
error distribution lies at $\lesssim 0.5$~mas~yr$^{-1}$,
which is much smaller than the $\sim 4$~mas~yr$^{-1}$
corresponding to UCAC4 \citep{Zac13}.
As mentioned before, the main advantage of using the
CdC-SF catalogue is the relatively high precision in
the proper motion values that will allow us to obtain
reliable kinematic memberships and to study with
unprecedented detail the proper motion distribution.
Thus, in this work we are using positions
and proper motions from the CdC-SF catalogue whereas
the photometric values are taken from UCAC4.
Table~\ref{tab:main} lists positions
(Equinox=J2000, Epoch=2000.0),
proper motions,
photometry and membership probabilities (calculated
in next section) for the full sample of stars we have
considered in this work.

\section{Membership determination}
\label{sec:members}

%%%%%%%%%%%%%%%%%%%%%%%%%%%%%%%%%%%%%%%%%%%%%%%%%%%%%%%%%%%%
\begin{table*}
\caption{Kinematic parameters for NGC~2548.}
\label{tab:duo}
\begin{tabular}{rcccccc}
\hline
& $N/N_t$ & $\rho$ & $\mu_\alpha\cos\delta$ & $\mu_\delta$
& $\sigma_{\mu_\alpha\cos\delta}$ & $\sigma_{\mu_\delta}$ \\
& & & (mas~yr$^{-1}$) & (mas~yr$^{-1}$) & 
      (mas~yr$^{-1}$) & (mas~yr$^{-1}$) \\
\hline
NGC~2548 & 0.61 & -0.26 & -1.55 & +3.16 & +3.09 &  +3.31 \\
Field    & 0.39 & -0.11 & -4.04 & -0.05 & +9.61 & +10.28 \\
\hline
\end{tabular}
\end{table*}
%%%%%%%%%%%%%%%%%%%%%%%%%%%%%%%%%%%%%%%%%%%%%%%%%%%%%%%%%%%%

The classical method to determine membership probabilities
from proper motion data was originally proposed by \citet{Vas58}
and later revised by \citet{San71} in the sense of fitting
techniques. The method assumes that the global distribution
of proper motions can be fitted as the sum of two components,
cluster and field, represented by two Gaussian functions.
Here we use the improved procedure developed by \citet{Cab85}
which uses a more robust and efficient iterative algorithm
for the estimation of the model parameters. Although the
method detects and rejects outliers, we have previously
removed those stars having too high proper motions (outside
the range $\pm 40$~mas~yr$^{-1}$) and/or stars with proper
motion errors higher than $5$~mas~yr$^{-1}$ in order to avoid
unrealistic fitting solutions. Moreover, to take into account
the possibility of asymmetries in the velocity space, we do
not use a circular Gaussian distribution for the cluster
but a Gaussian function with dispersions that can be
different in right ascension and declination.

%%%%%%%%%%%%%%%%%%%%%%%%%%%%%%%%%%%%%%%%%%%%%%%%%%%%%%%%%%%%
\begin{figure}
\includegraphics[width=\columnwidth]{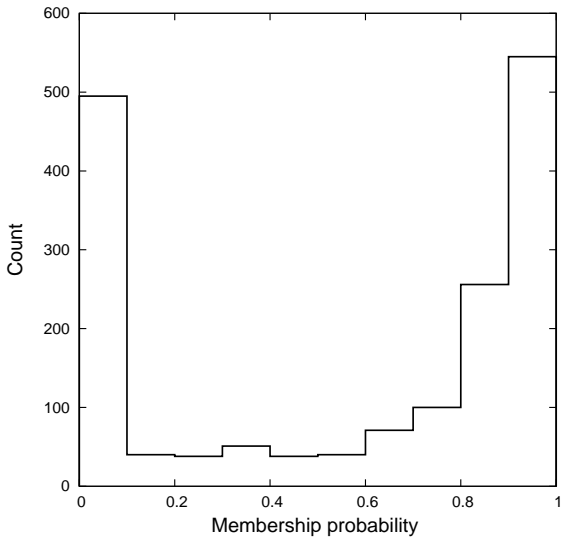}
\caption{Histogram of cluster membership probabilities
for stars in NGC~2548.}
\label{fig:prob}
\end{figure}
%%%%%%%%%%%%%%%%%%%%%%%%%%%%%%%%%%%%%%%%%%%%%%%%%%%%%%%%%%%%

The application of the algorithm to our data yielded
all star membership probabilities $p$ as well as
the final kinematic parameters for cluster and
field. Fig.~\ref{fig:prob} is the histogram of membership
probabilities where we can see that cluster and field stars
are clearly separated. There is a total of $1012$ members
($p>0.5$). The obtained kinematic parameters for both
cluster and field, including the fraction of stars
($N/N_t$), the correlation coefficient ($\rho$),
the proper motion in right ascension and declination
($\mu_\alpha\cos\delta$ and $\mu_\delta$) and their
corresponding dispersions ($\sigma_{\mu_\alpha\cos\delta}$
and $\sigma_{\mu_\delta}$), are summarized in 
Table~\ref{tab:duo}. We can see that cluster and field
are two populations that have been clearly differentiated
by the algorithm.

\section{Spatial and kinematic distribution of stars}
\label{sec:maps}

In order to calculate the spatial and kinematic density of
member stars in NGC~2548, we perform a direct estimation
using Kernel functions of the form
\begin{equation}
f(x,y) = \frac{1}{N h^2} \sum_{i=1}^N K \left(
\frac{x-x_i}{h} , \frac{y-y_i}{h} \right)
\end{equation}
being $x,y$ the spatial or kinematic variables and where a
Gaussian Kernel function $K$ is summed over the sample of $N$
members. The final density estimation depends considerably
on the value of the smoothing parameter $h$. To avoid either
very large $h$ values that over-smooth the distribution or
small values that produce noisy solution, we always use
the $h$ value such that the likelihood is maximum
\citep[see details in][]{Sil86,Cab90}.
For the spatial coordinates the optimal smoothing
parameter was $h=3.6$~arcmin whereas for the
proper motion space was $h=0.5$~mas~yr$^{-1}$.

%%%%%%%%%%%%%%%%%%%%%%%%%%%%%%%%%%%%%%%%%%%%%%%%%%%%%%%%%%%%
\begin{figure*}
\includegraphics[width=\columnwidth]{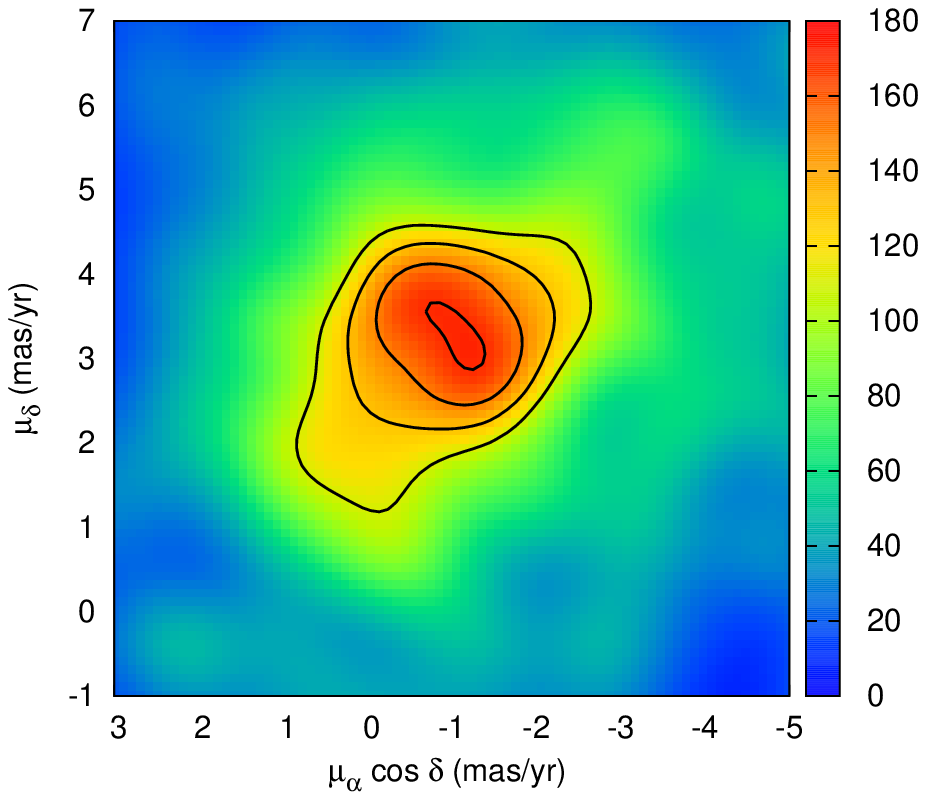}
\includegraphics[width=\columnwidth]{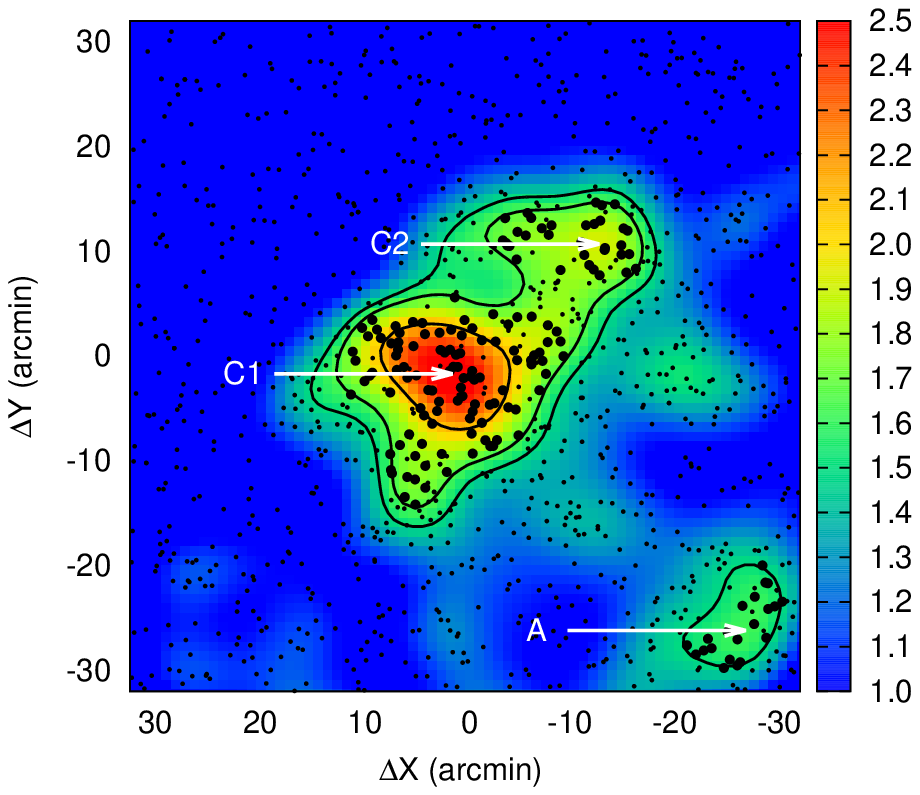}
\caption{Left panel: star density in the proper
motion space for the members of NGC~2548. Equally
spaced isocontours from $110$ to $170$ stars per
(mas~yr$^{-1}$)$^2$ are shown. Right panel: spatial
density of member stars in NGC~2548 (positions
are relative to the cluster's centre). The colour
scale goes from $1.0$ a $2.5$ stars~arcmin$^{-2}$ being
the average of the field $\sim0.8$ stars~arcmin$^{-2}$.
Isocontours are shown for $1.50$, $1.65$ and $2.10$
stars~arcmin$^{-2}$ corresponding to percentiles
$90$, $95$ and $99$\%, respectively. Arrows indicate
each one of the three identified maxima that we
associate with the three cores denoted by C1, C2
and A. Small black dots are cluster members and
bigger black points correspond to the stars selected
as belonging to these cores (see details in the text).}
\label{fig:mapas}
\end{figure*}
%%%%%%%%%%%%%%%%%%%%%%%%%%%%%%%%%%%%%%%%%%%%%%%%%%%%%%%%%%%%

The resulting density maps are shown in Fig.~\ref{fig:mapas}.
The distribution of proper motions (left panel in 
Fig.~\ref{fig:mapas}) is far from being the circular
distribution expected for typical open clusters.
The global shape of the proper motion distribution
is rather elongated, but interestingly the central 
part of the distribution is elongated along the direction
perpendicular to the global shape. On the other hand, the
spatial star distribution (right panel in Fig.~\ref{fig:mapas})
is quite complex. We can see an irregular main structure,
consisting of a relatively dense central core (marked as
C1) and a less dense secondary core (C2) clearly detached
from the main core. Additionally, there is a slight
overdensity in SW direction (marked as A) well separated
from the two central cores. These features (C1, C2 and A),
that were already described by \citet{Ber01}, stand out
clearly over the mean field star density, which is
$\sim0.8$ stars~arcmin$^{-2}$. The main core (C1) peaks
with $2.6$ stars~arcmin$^{-2}$ at $(\Delta X,\Delta Y) =
(+1.2,-1.7)$~arcmin relative to the assumed centre of
NGC~2548 \citep[taken from][]{Dia02}, whereas C2 shows
a peak of $1.9$ stars~arcmin$^{-2}$ at $(\Delta X,\Delta Y) =
(-12.8,+10.7)$~arcmin. The maximum value for A is $1.6$
stars~arcmin$^{-2}$ and is located at $(\Delta X,\Delta Y) =
(-26.8,-26.2)$~arcmin.

\begin{figure*}
\includegraphics[width=\columnwidth]{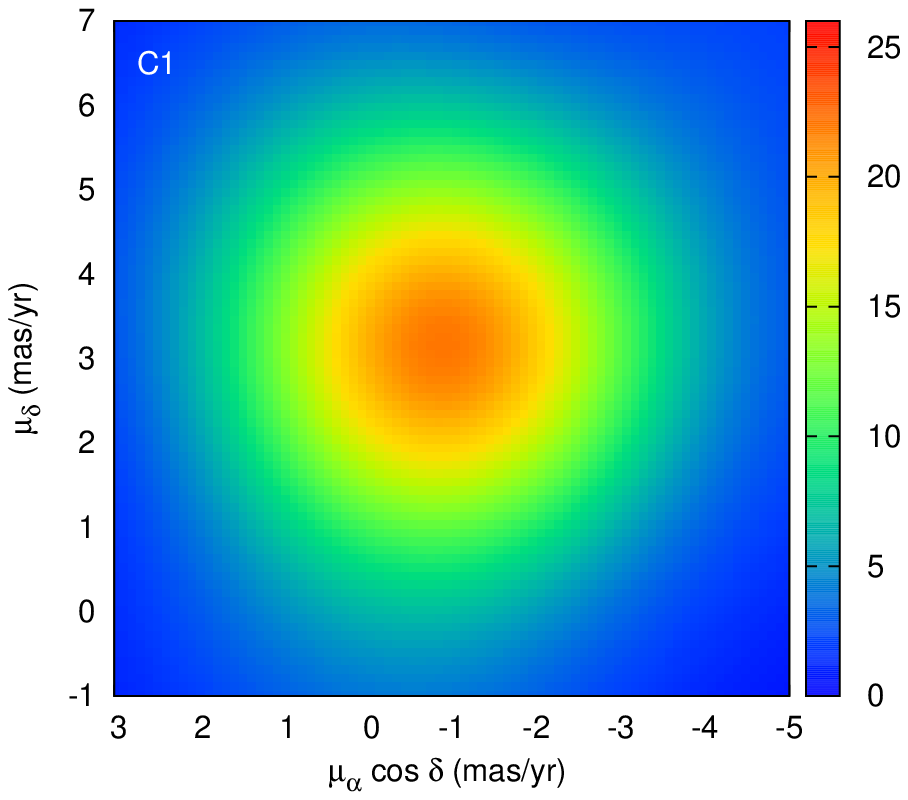}
\includegraphics[width=\columnwidth]{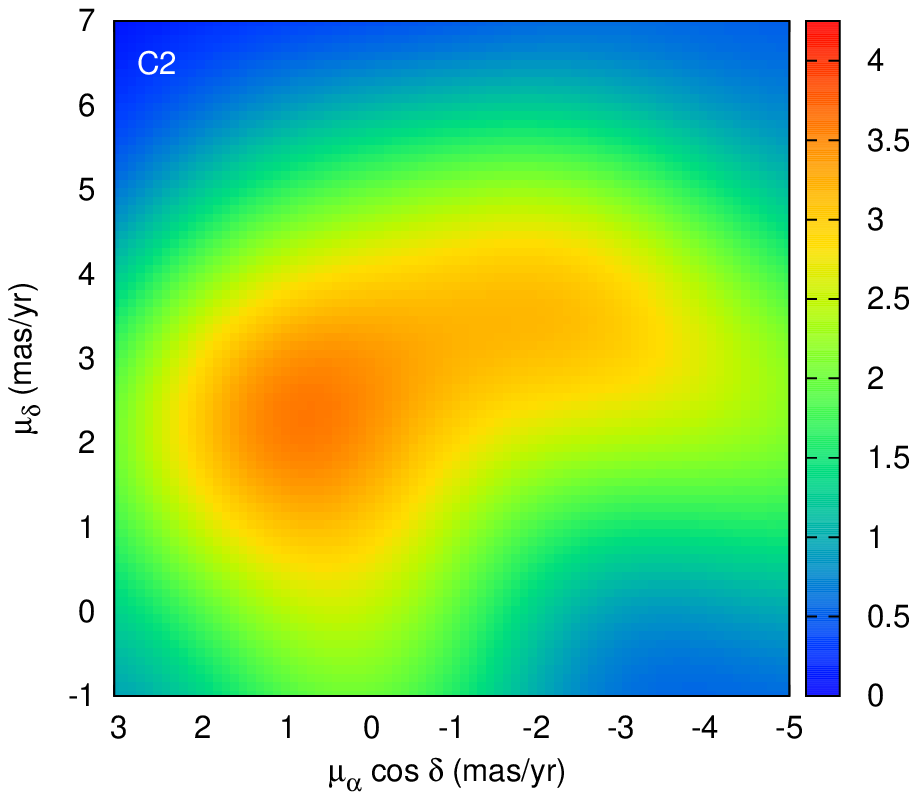}
\includegraphics[width=\columnwidth]{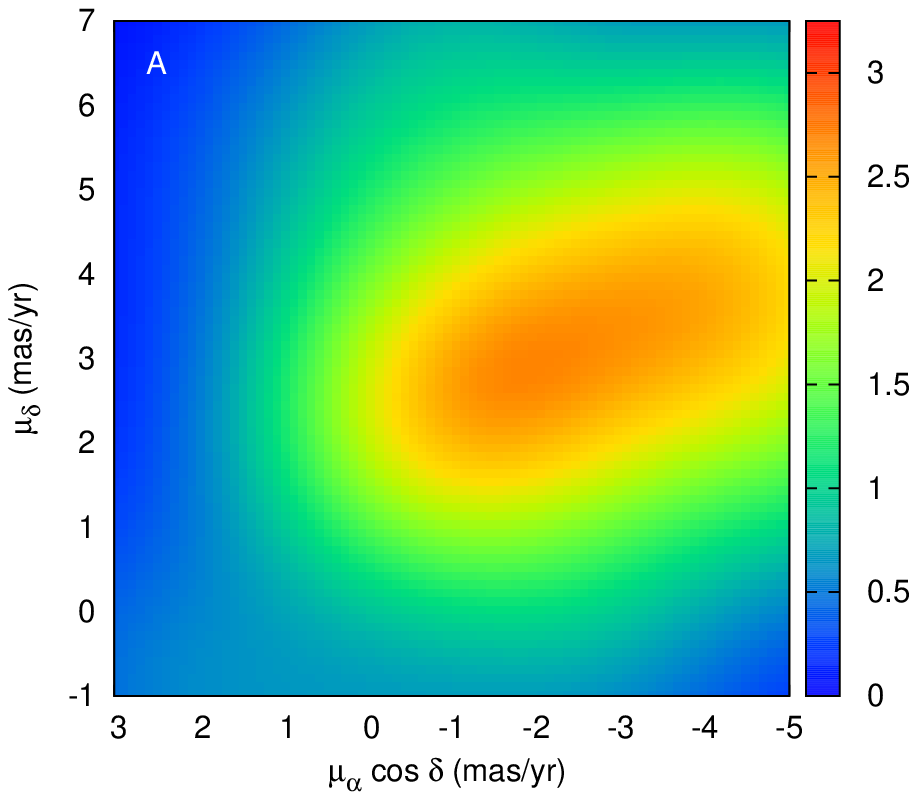}
\includegraphics[width=\columnwidth]{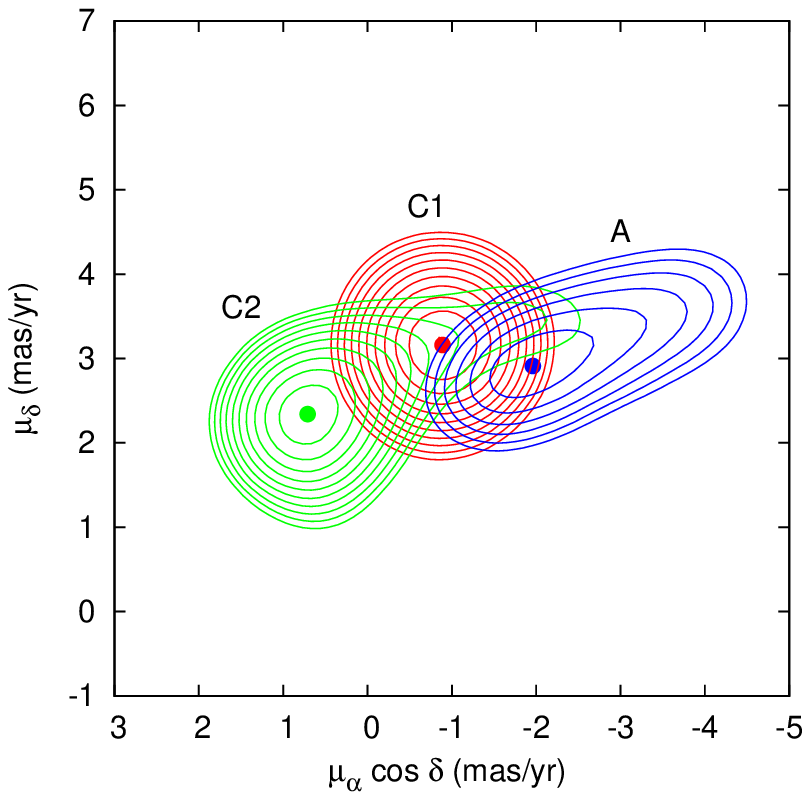}
\caption{Star density in the proper motion space for
the stars belonging to cores C1 (upper left panel),
C2 (upper right) and A (lower left). Equally spaced
isocontours above the $95$th percentile are shown
together in the lower right panel.}
\label{fig:cores}
\end{figure*}

In order to describe in more detail their kinematic
properties, we have selected the stars belonging to
each of the observed substructures. For the selection
we have taken into account the isocontours shown in
the right panel of Fig.~\ref{fig:mapas} that correspond
to the percentiles $90$, $95$ and $99$\% of the surface density
values. Then we select approximately the stars inside
these isocontours and associate them to each core based
on their proximity to the density peaks. For the two
main cores (C1 and C2) we use the $95$th percentile as
threshold whereas for A we use the $90$th. The resulting
selection is indicated in Fig.~\ref{fig:mapas} as black
points. Once selected, we recalculate the kinematic
density distribution for each core, following the same previous
procedure. The result is shown in Fig.~\ref{fig:cores}.
We note that each substructure (core) seen in the spatial
distribution of stars have a well differentiated counterpart
in the proper motion space. Proper motions of core C1
follow a somewhat regular, circular distribution with
the maximum located at $(\mu_\alpha\cos\delta , \mu_\delta) =
(-0.89,+3.16)$~mas~yr$^{-1}$. Cores C2 and A exhibit more
irregular distributions with less pronounced maxima at
$(\mu_\alpha\cos\delta , \mu_\delta) = (+0.71,+2.34)$~mas~yr$^{-1}$ and
$(\mu_\alpha\cos\delta , \mu_\delta) = (-1.96,+2.91)$~mas~yr$^{-1}$,
respectively. The C1-C2 and C1-A centroid separations 
in the vector-point diagram are $\sim1.8$
and $\sim1.1$~mas~yr$^{-1}$, respectively, above the typical
error for our proper motion values.

%%%%%%%%%%%%%%%%%%%%%%%%%%%%%%%%%%%%%%%%%%%%%%%%%%%%%%%%%%%%
%%%%%%%%%%%%%%%%%%%%%%%%%%%%%%%%%%%%%%%%%%%%%%%%%%%%%%%%%%%%

%%%%%%%%%%%%%%%%%%%%%%%%%%%%%%%%%%%%%%%%%%%%%%%%%%%%%%%%%%%%
\begin{table}
\caption{Derived properties for the three cores observed
in NGC~2548.}
\label{tab:cores}
\begin{tabular}{cccccc}
\hline
Core & $N$ & $\Delta X$ & $\Delta Y$
           & $\mu_\alpha\cos\delta$ &  $\mu_\delta$ \\
 & & (arcmin) & (arcmin) & (mas~yr$^{-1}$) & (mas~yr$^{-1}$) \\
\hline
C1 & 106 &  +1.2 & -1.7  & -0.89 & +3.16 \\
C2 &  34 & -12.8 & +10.7 & +0.71 & +2.34 \\
A  &  21 & -26.8 & -26.2 & -1.96 & +2.91  \\
\hline
\end{tabular}
\end{table}
%%%%%%%%%%%%%%%%%%%%%%%%%%%%%%%%%%%%%%%%%%%%%%%%%%%%%%%%%%%%

It is the first time, to our knowledge, that a clumpy
structure in the proper motion space is detected in
an open cluster, where each core
centroid can be unequivocally associated with a
corresponding spatial blob. In general terms,
the proper motions of these three substructures 
explain the observed kinematic distribution shown 
in left panel of Fig.~\ref{fig:mapas}: the combination 
of the proper motions of C1 and C2 produces the global
ellipsoidal shape whereas the combination with core A
generates the perpendicular elongation in the central 
part of the distribution.
Clump C1 is nearly spherical in both position
and proper motion space whereas C2 can be considered as
a small perturbation on the bulk of the cluster.
Clump A, on the other hand, is a weak overdensity
whose precise nature is uncertain (see discussion
later).
Table~\ref{tab:cores} summarizes the derived properties
for each core observed in NGC~2548, including the number
of stars $N$, their positions relative to the assumed
cluster centre and their representative proper motions.
Both positions and proper motions refer to the maxima
of their density distributions.
We have verified that the detection and shapes of these
substructures are not very sensitive to the exact value
of the chosen sampling radius and to the exact number of
assigned members.
Peak densities observed in right panel of
Fig.~\ref{fig:mapas} are $2-3$ times over
the background density. Interestingly,
membership probability values for stars
belonging to the clumps are, in general,
relatively high (see Table~\ref{tab:main}) 
so that if we set, for instance, a more restrictive 
membership threshold the total number of cluster
members decreases but clump members decrease in a
smaller proportion, in such way that the observed 
density peaks remain unchanged or even increase 
relative to the cluster background.

\subsection{Cluster photometry}

%%%%%%%%%%%%%%%%%%%%%%%%%%%%%%%%%%%%%%%%%%%%%%%%%%%%%%%%%%%%
\begin{figure}
\includegraphics[width=\columnwidth]{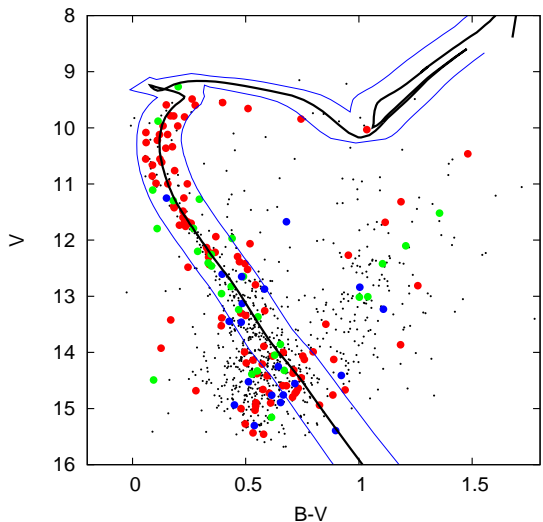}
\caption{Colour-magnitude diagram for all members of NGC~2548.
Red, green and blue circles mark stars belonging to cores C1,
C2 and A, respectively. As a reference, we have also plotted 
(solid black line) the Padova isochrone \citep{Bre12} of
age $500$~Myr, $E(B-V)=0.1$, solar metallicity, and distance
$780$~pc.
Blue lines define a band around this isochrone to allow for
an uncertainty of three times the mean error in magnitude
(0.020) and colour (0.032) and to consider the effects of
binarity.}
\label{fig:foto}
\end{figure}
%%%%%%%%%%%%%%%%%%%%%%%%%%%%%%%%%%%%%%%%%%%%%%%%%%%%%%%%%%%%

Figure~\ref{fig:foto} shows the $V$ vs. $B-V$ diagram for
all member stars in NGC~2548. Main sequence is clearly
visible and is fairly well fitted with an isochrone
corresponding to a slightly reddened ($E(B-V)=0.1$)
cluster of age of $500$~Myr located at $780$~pc. Most
stars associated to cores C1, C2 and A (coloured circles
in Fig.~\ref{fig:foto}) lie on the main sequence.
We see that stars belonging to the double core
C1-C2 (red and green circles) mostly follow the
isochrone upper-tail, while stars in core A appear fainter
and spread out over the lower-tail of the $(B,B-V)$ diagram.
The minimum magnitude for core A is $V\sim11.3$
with a median of $\sim14.3$, which is two magnitudes
higher than the C1-C2 median magnitude ($\sim12.4$).
Given that NGC~2548 is a practically non-extincted
cluster, this means that core A stars are on average
less luminous than C1-C2 stars.

\begin{figure}
\includegraphics[width=\columnwidth]{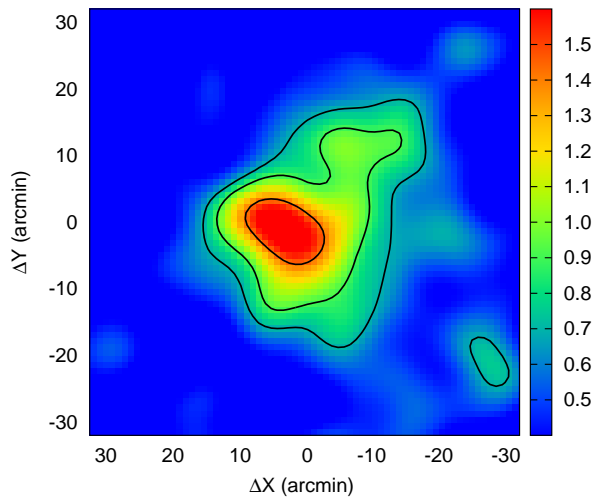}
\caption{Spatial density for stars that are
kinematic and also photometric members of the
cluster (see text). Colour scale goes from
$0.4$ to $1.6$ stars~arcmin$^{-2}$ and
isocontours are shown for percentiles
$90$, $95$ and $99$\%.}
\label{fig:mapaphoto}
\end{figure}

Stars plotted in Fig.~\ref{fig:foto} are all the
kinematically selected members of NGC~2548, although
we see that a few stars have magnitude and colour values
such that they are not cluster members from a photometric
point of view. In this work we have left cluster memberships
unaltered because the issue of kinematic versus photometric
(or combined kinematic-photometric) cluster memberships,
their reliabilities and possible effects on the derived 
cluster properties is out of the scope of this paper.
However, in order to evaluate the robustness of our 
results we performed some additional tests.
We defined a region around the considered isochrone
in the B-V colour-magnitude diagram (delimited by blue
lines in Fig.~\ref{fig:foto}), then we selected as
kinematic {\it and} photometric members those stars
lying inside this area and we recalculated the spatial
density of stars as before. The result is shown in
Fig.~\ref{fig:mapaphoto} and Table~\ref{tab:fotocore}
summarizes some core properties for the cases of
only kinematically selected members and also
photometrically selected members in the B-V
diagram.\footnote{We did several tests considering
different photometric bands and the results remain
qualitatively similar. Here we show the B-V case
as an example, a rigorous photometric analysis would
involve several photometric bands simultaneously.}
The number of cluster members and therefore the
mean star density decrease but, as it happened
when we applied a more restrictive kinematic membership
selection threshold, core members decrease in a smaller
proportion and their density peaks increase relative to
the background. As before, core A is seen above the
$90$th percentile but as a weak overdensity with few
members.

%%%%%%%%%%%%%%%%%%%%%%%%%%%%%%%%%%%%%%%%%%%%%%%%%%%%%%%%%%%%
%%%%%%%%%%%%%%%%%%%%%%%%%%%%%%%%%%%%%%%%%%%%%%%%%%%%%%%%%%%%

%%%%%%%%%%%%%%%%%%%%%%%%%%%%%%%%%%%%%%%%%%%%%%%%%%%%%%%%%%%%
\begin{table}
\caption{Comparison of core properties for
kinematically selected and kinematically plus photometrically
selected cluster members (density unit is stars~arcmin$^{-2}$).}
\label{tab:fotocore}
\begin{tabular}{lcc}
\hline
         & Kinematic & Photometric \\
Property & members   & members \\
\hline
Cluster members  & 1012 & 438 \\
Mean cluster density & 0.8 & 0.3 \\
Core stars (C1, C2, A) & 106 , 34 , 21 & 68 , 18 , 9 \\
Percentil (Q99, Q95, Q90) & 2.1 , 1.7 , 1.5 & 1.3 , 0.9 , 0.7 \\
Peak density (C1, C2, A) & 2.6 , 1.9 , 1.6 & 1.8 , 1.0 , 0.8 \\
\hline
\end{tabular}
\end{table}
%%%%%%%%%%%%%%%%%%%%%%%%%%%%%%%%%%%%%%%%%%%%%%%%%%%%%%%%%%%%

\section{Discussion}
\label{sec:discussion}

Our main result is the finding that each substructure
observed in the spatial distribution of stars in NGC~2548
has its well differentiated counterpart in the proper
motion space, which provides the first {\it kinematic}
evidence of cluster disruption occurring in this stellar
system. As far as we know, this is the first detection
of a clumpy structure in the proper motion space in any
star cluster.
As discussed by \citet{Ber01}, the spatial clumpy structure
observed in NGC~2548 is likely a consequence of its interaction
with the Galactic gravitational field. As clusters move on their
orbits through the Milky Way, they will pass several times through
the Galactic disk and they may become morphologically disturbed
or even destroyed due to the Galactic tidal forces \citep{Ter87}.
Successive passes through the Galactic plane can speed up the
destruction of clusters, although the ease with which they can be
destroyed depends on variables such as the Initial Mass Function
or the initial virial state, and the expected disruption time-scale
varies from less than $\sim100$~Myr up to the order of $\sim1$~Gyr
\citep{Fue97}. Other possible disruption mechanisms, such as close
encounters with giant molecular clouds, do not seem to be as effective
as Galactic tides in disrupting open cluster \citep{Ter87,Gie06}.

%%%%%%%%%%%%%%%%%%%%%%%%%%%%%%%%%%%%%%%%%%%%%%%%%%%%%%%%%%%%
\begin{figure}
\includegraphics[width=\columnwidth]{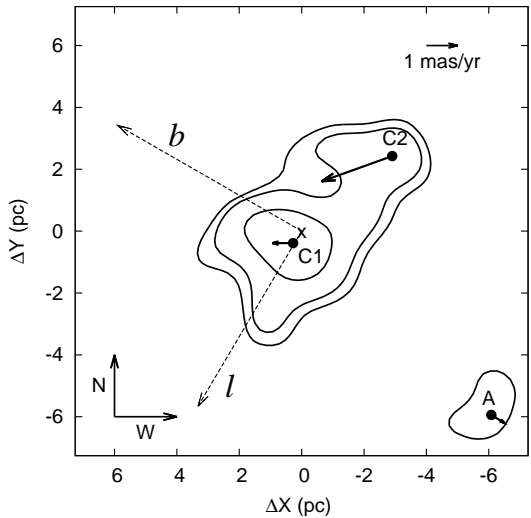}
\caption{Simplified version of Fig.~\ref{fig:mapas}
(right panel) showing the higher density isocontours
and positions of cores C1, C2 and A. Solid arrows
represent the proper motions of the cores relative
to the global proper motion of the cluster derived
in Section~\ref{sec:members}. Reference arrow in
upper right corner has a length of $1$~mas~yr$^{-1}$.
Dashed arrows indicate axes parallel ($l$) and
perpendicular ($b$) to the Galactic plane,
respectively.}
\label{fig:paladiscu}
\end{figure}
%%%%%%%%%%%%%%%%%%%%%%%%%%%%%%%%%%%%%%%%%%%%%%%%%%%%%%%%%%%%

The relative spatial configuration of the three internal
cores in NGC~2548 suggests a predominant interaction with
Galactic forces. Figure~\ref{fig:paladiscu} shows the
spatial distribution of the cores, also indicating
their proper motions relative to the cluster centroid
and the directions perpendicular and parallel to the
Galactic plane. The stretching of C1-C2 is nearly
parallel to the Galactic plane, which is consistent
with the flattening in the same direction predicted
in numerical simulations \citep{Ter87}. On the other hand,
clump A is located relatively far from the main
cores at $\sim8.5$~pc in the direction of the
Galactic plane. This was interpreted by \citet{Ber01}
as A being a remnant of the last disk-shocking
of NGC~2548. Note that A's proper motion relative
to the cluster centroid (black arrows in
Fig.~\ref{fig:paladiscu}) goes towards the
Galactic plane and away from the rest of the
cluster.

Nevertheless, it must be pointed out that it is
not possible to be sure about the real nature of
clump A. Given that it is a weak overdensity
with few members and it seems to be detached
from the main cluster (cores C1-C2), a random
concentration of field stars is also a plausible
explanation for the origin of this clump.
However, we should note that we are not
estimating spatial overdensities from the full
star sample without any rigorous membership 
analysis \citep[as in][]{Ber01}, but after
a reliable kinematic membership assignment.
The fact that clump A is inside our estimated 
cluster radius together with the fact that
its presence is robust against the kinematic
pruning of the data point toward a real star
clump moving away from the central part.
There is still the alternative possibility
that clump A is a small spatial overdensity of
stars but lying at some distance
behind NGC~2548. This alternative is supported
by the, on average, fainter magnitudes of
their stars but, in this case, the kinematic
similarities have to be interpreted as
happening by chance. If we take as true
that A is a clump that has been separated
from the original cluster by tidal forces,
then the kinematic solutions are consistent
because A goes away from the rest of the
cluster due to its last pass through the
Galactic disk.
The challenge in this case is to understand
why the core A is on average less luminous
than C1-C2. There must have been some
mass-segregation episode either associated
with the dynamical disruption itself or with
a primordial mass-segregated pattern.

There are many open clusters showing elongations or other shape
distortions \citep{Che04}. For a very young open cluster, its
morphological structure is likely determined by the initial
conditions in the parental cloud \citep{San09}; but for
dynamically evolved clusters, external tidal perturbations
would become increasingly important. In fact, cluster
properties such as size or flattening seem to correlate
with distances to the Galactic centre and plane probably
because these variables define the degree of tidal stress
\citep{Che04,Bon10}. It would be necessary to extend
this type of analysis to large samples of disrupting
star cluster candidates \citep[as, for instance, those
in][]{Bic01}.
GAIA data will be particularly useful for this because it
will measure very precise proper motion values even for
low-mass members in open clusters. In this work, we were
able to detect kinematic substructures because of our 
relatively precise proper motions (mean error
$\sim 1.0$~mas~yr$^{-1}$, four times better than UCAC4),
but GAIA expected mean errors are $\sim0.08$~mas~yr$^{-1}$
(about ten times smaller than ours) or even
$\sim0.02$~mas~yr$^{-1}$ for bright stars with $G<15$
\citep{Lur14}. Moreover, GAIA will also provide radial
velocities that make it possible to derive reliable
three-dimensional space velocities. This will 
allow to perform detailed statistical studies
to better understand
the cluster destruction processes and how they
depend on cluster environmental properties such as
Galactic location and orbit.

\section{Conclusions}
\label{sec:conclusion}

We have used precise astrometric data from the CdC-SF catalogue
to analyse the spatial and kinematic structure of the open
cluster NGC~2548. From the spatial distribution of
reliable cluster members
we confirm its previously reported fragmented structure
consisting of three separated cores (denoted as C1, C2 and A)
but, for the first time, we also distinguish the corresponding
blobs in the proper motion space. The two main cores C1 and C2
are aligned nearly parallel to the Galactic plane whereas the
fainter and less dense core A moves toward the plane and
appears to be detaching itself from the rest of the cluster.
We present the derived core properties in Table~\ref{tab:cores}.
This clumpy spatial and kinematical structure observed in
NGC~2548 could be consequence of the Galactic gravitational
field, in particular due to its last crossing through the Galactic
plane.

\section*{Acknowledgements}
We thank the referee for his/her comments which improved this paper.
We acknowledge financial support from Ministerio de Econom\'{\i}a y
Competitividad of Spain and FEDER funds through grant AYA2013-40611-P.
NS has received partial financial support from Fundaci\'on S\'eneca 
de la Regi\'on de Murcia (19782/PI/2014) and Ministerio de
Econom\'{\i}a y Competitividad of Spain (FIS-2015-32456-P).

\bsp
\label{lastpage}
\end{document}